\documentclass[conference,a4paper]{IEEEtran}

\usepackage{cite}
\usepackage{graphicx,color,epsfig,rotating}
\usepackage{amsfonts,amsmath,amssymb,bbm,bm}
\usepackage{algorithm,algorithmic}
\usepackage{subfigure}
\usepackage{amsmath}
\usepackage{cite}
\usepackage{mdwtab}
\usepackage{placeins}
\usepackage{psfrag, graphicx}
\usepackage[latin1]{inputenc}
\usepackage{amssymb}
\usepackage{multirow}

\ifodd 1

\else

\fi

\newtheorem{theorem}{Theorem}%[section]
\long\def\comment#1{}

\newfont{\bbb}{msbm10 scaled 700}

\newfont{\bb}{msbm10 scaled 1100}

\newcommand{\Nc}{{\cal N}}

\newcommand{\be}{\begin{equation}}
\newcommand{\ee}{\end{equation}}
\newcommand{\bea}{\begin{eqnarray}}
\newcommand{\eea}{\end{eqnarray}}

\begin{document}

\sloppy

%% Paper Title
%% You can use linebreaks \\ within to get better formatting as
%% desired.

\title{A New Design of Private Information Retrieval for Storage Constrained Databases}

%\title{Storage Constrained Private Information Retrieval -- Should PIR meet Coded Caching?}

\author{
    \IEEEauthorblockN{ Nicholas Woolsey,
		Rong-Rong Chen, and Mingyue Ji }
	\IEEEauthorblockA{Department of Electrical and Computer Engineering, University of Utah\\
		Salt Lake City, UT, USA\\
		Email: \{nicholas.woolsey@utah.edu,
		 rchen@ece.utah.edu,
		mingyue.ji@utah.edu\}}

}

%% Create the title:
\maketitle

%\vspace{-5cm}

\begin{abstract}
Private information retrieval (PIR) allows a user to download one of $K$ messages from $N$ databases without revealing to any database which of the $K$ messages is being downloaded. In general, the databases can be storage constrained where each database can only store up to $\mu K L$ bits where $\frac{1}{N} \leq \mu \leq 1$ and $L$ is the size of each message in bits. Let $t= \mu N$, a recent work showed that the capacity of Storage Constrained PIR (SC-PIR) is $\left( 1+ \frac{1}{t} + \frac{1}{t^2} + \cdots + \frac{1}{t^{K-1}} \right)^{-1}$, which is achieved by a storage placement scheme inspired by the content placement scheme in the literature of coded caching and the original PIR scheme.
Not surprisingly, this achievable scheme requires that each message is $L = {N \choose t}t^K$ bits in length, which can be impractical. In this paper, without trying to make the connection between SC-PIR and coded caching problems, based on a general connection between the Full Storage PIR (FS-PIR) problem ($\mu = 1$)  and SC-PIR problem, we propose a new SC-PIR design idea using novel storage placement schemes. The proposed schemes %that
significantly reduce the message size requirement while still meeting the capacity of SC-PIR.
In particular, the proposed SC-PIR schemes require the size of each file  to be only $L = Nt^{K-1}$ compared to the state-of-the-art $L = {N \choose t}t^K$. Hence, we conclude that PIR may not meet coded caching when the size of $L$ is constrained.
%where %$t\triangleq \mu N$. %However, %An achievable storage constrained PIR scheme was developed which performs at capacity. However,
%A fundamental connection between the full storage PIR (FS-PIR) scheme ($\mu = 1$) and the  storage constrained PIR (SC-PIR) scheme is established so that  a SC-PIR scheme can be derived from any  FS-PIR scheme. Using a newly designed  FS-PIR scheme, we obtain
%for a better design. %, which is the state-of-the-art in the literature. %$L = \Omega\left( N^K \right)$ bits in length while achieving capacity. \mj{[Is it the same as Sun and Jafar's scheme?]}
%Here, the proposed storage constrained PIR schemes can be induced from any full storage schemes
%To do this, we demonstrate the storage constrained PIR problem can be solved using an multiple adapted full storage ($\mu = 1$) PIR scheme. We use two full storage PIR schemes, the state-of-the-art scheme and a new scheme, to develop two storage constrained PIR schemes which meet capacity while only requiring that each file is $L ~ = O\left( N^K \right)$ bits in length.

\end{abstract}
%%%%%%%%%%%%%%%%%%%%%%%%%%%%%%%%%%%%%%%%%%%%%%%
%%%%%%%%%%%%%%%%%%%%%%%%%%%%%%%%%%%%%%%%%%%%%%%
\section{Introduction}
\label{sec: Intro}
Recent works have taken an information theoretic approach to solve the private information retrieval (PIR) problem \cite{sun2017capacity,attia2018capacity} originally introduced by Chor \textit{et al.} \cite{chor1995private, chor1998private}. In the PIR problem, a user desires to privately download one of $K$ messages from $N$ non-colluding databases. In this context, privacy means that the identity of the message desired by the user is not revealed to any database.
%does not know the identity %has no insight into which
%of the messages the user desires.
Ensuring privacy relies on the %simple
concept that a user will request sub-messages from all $K$ messages as opposed to just the message that the user desires. To efficiently download the desired message, the user strategically generates database queries that utilize undesired but downloaded sub-messages for coding opportunities.  The rate %$R$
of a PIR scheme is defined  as the ratio of desired bits, $L$, or the size of each message, to the total number of downloaded bits, $D$. The capacity $C$ (optimal rate) is defined as the maximum achievable rate.

Previously, Sun and Jafar \cite{sun2017capacity} derived the capacity of the Full Storage PIR (FS-PIR) problem where a user privately downloads one of $K$ messages from $N$ databases that each stores all $K$ messages. In this case, the capacity %optimal rate \mj{[Should we say capacity here?]}
is $C = \left( 1+ \frac{1}{N} + \frac{1}{N^2} + \cdots + \frac{1}{N^{K-1}} \right)^{-1}$, which was achieved by a PIR scheme requiring $L = N^{K}$. %\mj{[Say a word here about the subpacketization level of Sun and Jafar's scheme]}
This result was further generalized by Attia \textit{et al.} \cite{attia2018capacity} for the Storage Constrained PIR (SC-PIR) problem where each database can only store $\mu K L$ uncoded bits where $\frac{1}{N} \leq \mu \leq 1$. %They demonstrated
In this case, both a storage placement scheme and a PIR scheme (querying and decoding)  need to be designed.
Let $t=\mu N$, the capacity of SC-PIR is $\left( 1+ \frac{1}{t} + \frac{1}{t^2} + \cdots + \frac{1}{t^{K-1}} \right)^{-1}$ under an uncoded storage placement constraint and was achieved by a storage placement scheme inspired by the coded caching problem \cite{maddah2014fundamental} and a PIR scheme based on \cite{sun2017capacity}.
%\cite{attia2018capacity}. Inspired by the coded caching problem \cite{maddah2014fundamental}, the authors of \cite{attia2018capacity} developed a scheme which achieves this optimal rate for the storage constrained PIR (SC-PIR) problem. However,
One of the limitations of this scheme is the requirement of a large message size, $L = {N \choose t}t^K$ \cite{attia2018capacity}, which is due to the fact that the storage placement is designed based on the cache placement in coded caching problem \cite{maddah2014fundamental}. %Similar to coded caching schemes \cite{maddah2014fundamental}, %,ji2016fundamental,yan2017placement,woolsey2017towards},
Hence, the proposed PIR scheme of \cite{attia2018capacity} can be impractical for a large number of databases. % since $L = O\left(e^N\right)$. \mj{[Need to add a discussion on Ulukus's decentralized paper.]}
This achievable scheme was generalized to the decentralized storage placement in \cite{wei2018capacity}. Furthermore, Tian {\em et al.} \cite{tian2018PIR} use Shannon theoretic approach to analyze the SC-PIR problem for the canonical case of $K=2$ and $N=2$ and proposed the optimal linear scheme. More interestingly, they also showed that non-linear scheme can use less storage than the optimal linear scheme.

In this paper, we aim to find SC-PIR schemes that achieve the capacity of SC-PIR while  %with optimal %or near-optimal
%rate %\mj{[WHY near optimal?]}
%which requires
requiring a significantly smaller message size $L$. %\mj{Interestingly, we find that  there exists achievable schemes that are much simpler than those inspired by coded caching schemes. [This sentence needs to be written]}
In order to achieve this goal, for the storage placement, we abandon the idea of using the cache placement of coded caching problem and design it from scratch. In fact, our proposed SC-PIR schemes achieve the capacity and require only $L = Nt^{K-1}$, which is significantly less than $L = {N \choose t}t^K$ in \cite{attia2018capacity}. % to be polynomial with respect to $N$, or more specifically, $L = O\left( N^K \right)$.
More specifically, our contributions are as follows.

\paragraph*{Our Contributions}
\begin{enumerate}
\item We provide a general design methodology for the SC-PIR problem by establishing a generic connection between the FS-PIR and SC-PIR problems. Based on this connection, a SC-PIR scheme can be readily designed from any given FS-PIR scheme.
\item We propose a simple storage placement when $\frac{N}{t}$ is an integer. By adopting the achievable scheme based on \cite{sun2017capacity}, the capacity of SC-PIR can be achieved and $L = Nt^{K-1}$. This serves as a base case for the more general scenario when $\frac{N}{t}$ is not an integer. %\mj{[I have a question here, if your proposed SC-PIR scheme is used, do we get $L=Nt^{K-2}$?]}
\item When $\frac{N}{t}$ is not an integer, we propose a novel storage placement, which in conjunction with the FS-PIR scheme of \cite{sun2017optimal}, achieves the capacity of SC-PIR and only requires $L = Nt^{K-1}$. The key to the reduction in $L$ is achieved using the proposed novel storage placement.
%\item r we design two novel storage placement schemes for different parameter regimes, which are not designed from the idea of coded caching as in \cite{attia2018capacity,wei2018capacity}. The proposed new storage placement scheme only requires ?
%\item We provide a general design methodology for the storage constraint PIR problem by establishing a general connection between the FS-PIR and SC-PIR problems. Based on this connection, a SC-PIR scheme can be readily designed from any given FS-PIR scheme.
%Our proposed schemes are based on a general connection between FS-PIR and SC-PIR problems.
%In order to prove the achievable rate,
%Based on the storage placement, we establish a fundamental connection between FS-PIR and SC-PIR problems. Based on this connection, a SC-PIR scheme can be readily designed from any given FS-PIR scheme. Specifically, when adapting the FS-PIR scheme in Sun and Jafar\cite{sun2017capacity}, this finding leads to a simple SC-PIR scheme with optimal rate while requiring $L =  Nt^{K-1}$, under the assumption that  $\frac{N}{t} = \frac{1}{\mu} \in \mathbb{Z}^+$. To alleviate this assumption, we develop a new  FS-PIR scheme and the resulting SC-PIR scheme achieves the optimal rate while requiring $L = Nt^{K-1}$ for any $t \in [N]$ \mj{[Did we define the meaning of $[N]$?]}.
\item %Finally, using our SC-PIR design methodology, we define a set of sufficient conditions which yields a capacity achieving SC-PIR scheme.
We present  a set of sufficient conditions under which the proposed SC-PIR schemes are capacity-achieving.
\end{enumerate}
%\textit{Contributions}: We establish a fundamental connection between FS-PIR and SC-PIR problems. Based on this connection, a SC-PIR scheme can be readily designed from any given FS-PIR scheme. Specifically, when adapting the FS-PIR scheme in Sun and Jafar\cite{sun2017capacity}, this finding leads to a simple SC-PIR scheme with optimal rate while requiring $L =  Nt^{K-1}$, under the assumption that  $\frac{N}{t} = \frac{1}{\mu} \in \mathbb{Z}^+$. To alleviate this assumption, we develop a new  FS-PIR scheme and the resulting SC-PIR scheme achieves the optimal rate while requiring $L = Nt^{K-1}$ for any $t \in [N]$.
\paragraph*{Notation Convention}
%We use the following notation convention.
%Calligraphic symbols denote sets,
%bold symbols denote vectors,
%and sans-serif symbols denote system parameters.
We use $|\cdot|$ to represent the cardinality of a set or the length of a vector %;
%$[a:b]:=\left\{ a,a+1,\ldots,b\right\}$
and $[n] := [1,2,\ldots,n]$. %;
%$\oplus$ represents bit-wise XOR.

\section{Problem Formulation}
\label{sec: problem}
%\mj{[Did we say the databases cannot talk to each other? We need to write more in this section. Look at Tandon's paper!!]}
There are $K$ independent messages, $W_1 ,\ldots , W_K$, each of size $L$ bits. The messages are collectively stored in an uncoded fashion among $N$ non-colluding databases that each has a storage capacity of $\mu K L$ bits, where $\frac{1}{N} \leq \mu \leq 1$.  We define $Z_n$ %to represent
as the storage contents of database $n\in [N]$.
Also, we define
$t \triangleq \mu N$
as the average number of times each bit of the messages is stored among the databases.
A user makes a request $W_k$ and sends a query $Q_n^{[k]}$, which is independent of the messages, to each database $n\in [N]$ which then sends an answer $A_n^{[k]}$ such that
\be
\label{eq: PIR 1}
H(A_n^{[k]} | Z_n , Q_n^{[k]}) = 0, \quad \forall k \in [K].
\ee
Furthermore, given the answers from all the databases, the user must be able to recover the requested message with a small probability of error. Therefore,
\be
\label{eq: PIR 2}
H(W_k | A_1^{[k]}, \ldots , A_n^{[k]} , Q_1^{[k]} , \ldots , Q_n^{[k]}) = 0.
\ee

The user generates queries in a manner to ensure privacy such that no database has insight into which message  the user desires, {\em i.e.}, % \mj{[Define this in terms of entropy and mutual information.]} More rigorously,
%This means that the queries must follow the same distribution regardless of what file the user is requesting. For all $n \in [N]$ and all $i,j \in [K]$, $i \neq j$
%\begin{align}
%(Q_n | \theta = i) \sim (Q_n | \theta = j).
%&(Q_n^{[i]}, A_n^{[i]}, W_1, \ldots, W_K, Z_1, \ldots, Z_K) \notag\\
%&\sim (Q_n^{[j]}, A_n^{[j]}, W_1, \ldots, W_K, Z_1, \ldots, Z_K).
%\end{align}
\be
\label{eq: PIR 3}
I(k; Q_n^{[k]}, A_n^{[k]}, W_1, \ldots, W_K, Z_1, \ldots, Z_N) = 0.
\ee
Let D be the total number of downloaded bits. Given  $\mu$, we say that a pair $(D,L)$ is achievable  if there exists a SC-PIR scheme with rate $R=L/D$ that satisfies  (\ref{eq: PIR 1})-(\ref{eq: PIR 3}). The SC-PIR capacity is defined as
\be
C^*(\mu) = \max\{R: (D,L) \text{ is achievable}\}.
\ee
%The rate of a PIR scheme is  defined as
%\be
%R \triangleq \frac{L}{D},
%\ee
%which is the ratio of number of desired bits $L$ to the total number of downloaded bits $D$. \mj{[We need to define both rate and capacity, which are used arbitrarily in our paper.]}

%\section{A connection between FS-PIR and SC-PIR}
\section{The proposed SC-PIR scheme when $\frac{N}{t} \in \mathbb{Z}^+$}
\label{sec:connection}
In order to present the proposed scheme, we need to establish a connection between FS-PIR and SC-PIR problems.
%we establish a fundamental connection between FS-PIR and SC-PIR problems.
This connection is vital to reduce the required minimum size of messages from ${ N \choose t } t^K $, as in the state-of-the-art scheme of \cite{attia2018capacity}, to $Nt^{K-1}$ without affecting the optimal rate. We show that  an achievable SC-PIR scheme can be derived from any general achievable scheme for the FS-PIR problem. %In particular, the scheme of Sun and Jafar \cite{sun2017capacity} is adapted to obtain a new  SC-PIR scheme. \mj{[Why do you say it?]}
Hence, by using the proposed storage placement,  the achievable scheme in \cite{sun2017capacity} can be used to obtain a new  SC-PIR scheme. To illustrate our idea, we first present an example as follows.

\subsection{A Storage Constrained PIR Example when $\frac{N}{t} \in \mathbb{Z}^+$}
\label{sec: ex1}
Consider $N=4$ databases labeled as DB$1$ through DB$4$. Collectively the databases store $K=3$ messages, denoted by $A$, $B$ and $C$. Each message is comprised of $L=16$ bits. %as follows
\subsubsection{Storage placement scheme}
We split each message as follows.
\begin{align}
A &= \left\{ a_i^j : i \in [2], j \in [8] \right\} \\
B &= \left\{ b_i^j : i \in [2], j \in [8] \right\} \\
C &= \left\{ c_i^j : i \in [2], j \in [8] \right\}.
\end{align}
Each database has the storage capacity of up to $24$ bits, or half of all $3$ messages $\left(\mu = \frac{1}{2}\right)$. The storage contents of the databases are defined to be
\begin{align}
Z_1 = Z_2 &= \left\{ a_1^j : j\in [8] \right\} \cup \left\{ b_1^j : j\in [8] \right\} \cup\left\{ c_1^j : j\in [8] \right\} \\
Z_3 = Z_4 &= \left\{ a_2^j : j\in [8] \right\} \cup \left\{ b_2^j : j\in [8] \right\} \cup\left\{ c_2^j : j\in [8] \right\}.
\end{align}
\subsubsection{PIR Scheme}
Each database stores $8$ out of $16$ bits of each message. Databases $1$ and $2$ have the same storage contents, but do not have any storage contents in common with databases $3$ and $4$. Likewise, databases $3$ and $4$ have the same storage contents.
 In this way, we essentially reduce a SC-PIR problem into two independent FS-PIR problems; one consists of databases 1 and 2, and the other  consists of databases 3 and 4. Subsequently, we can simply adopt the achievable FS-PIR scheme of \cite{sun2017capacity} to generate the queries for each pair of the databases separately. The queries of a user that desires message A are shown in Table \ref{table: ex_1}.

\begin{table}[h!]
\vspace{-0.4cm}
\normalsize
\centering
{\small
\caption{ Storage Constrained PIR, $N=4$, $K=3$, $\mu = \frac{1}{2}$}
\vspace{-0.2cm}
 \label{table: ex_1}
\begin{tabular}{ |>{\centering}m{1.7cm}|>{\centering}m{1.7cm}||>{\centering}m{1.7cm}|>{\centering}m{1.7cm}| }
 \hline
 DB$1$ & DB$2$ & DB$3$ & DB$4$ \\
 \hline
 $a_1^5 \quad b_1^8 \quad c_1^6$ & $a_1^1 \quad b_1^3 \quad c_1^1$ & $a_2^5 \quad b_2^7 \quad c_2^4$ & $a_2^2 \quad b_2^6 \quad c_2^2$ \\
 \hline
 $a_1^6+b_1^3$ & $a_1^3+b_1^8$ & $a_2^1+b_2^6$ & $a_2^7+b_2^7$ \\
 $a_1^7+c_1^1$ & $a_1^8+c_1^6$ & $a_2^6+c_2^2$ & $a_2^8+c_2^4$ \\
 $b_1^6+c_1^5$ & $b_1^7+c_1^3$ & $b_2^3+c_2^6$ & $b_2^8+c_2^7$ \\
 \hline
 $a_1^2+b_1^7+c_1^3$ & $a_1^4+b_1^6+c_1^5$ & $a_2^3+b_2^8+c_2^7$ & $a_2^4+b_2^3+c_2^6$ \\
 \hline
\end{tabular}
}
 \end{table}

 \subsubsection{Achievable Rate}
 The total number of downloaded bits is $D = 28$.   Thus, we have for this scheme {$R=\frac{L}{D}=\frac{16}{28}=\frac{4}{7}$}, which achieves the capacity %\mj{[Now you use capacity, before we use optimal rate!!!]}
 of  $(1+\frac{1}{t}+\frac{1}{t^2})^{-1}=(1+\frac{1}{2}+\frac{1}{2^2})^{-1}=\frac{4}{7}$. Compared to the SC-PIR scheme of \cite{attia2018capacity} that requires $L = {N \choose t}t^K={4 \choose 2}2^3=48$ bits, the proposed SC-PIR requires only $L=16$ bits.

%Assume that a user desires message $A$, but does not want to reveal this information to the databases. The user generates a query for each database which can be found in Table \ref{table: 1}.
%Given that the user receives the requested bits, there are a few key observations of Table \ref{table: 1} which demonstrate that the user is able to decode $A$ and each node is unaware which message is being requested.

 \subsubsection{Privacy Constraint}
Privacy is ensured since the FS-PIR scheme of \cite{sun2017capacity} is used to privately download half of message $A$ from DB$1$ and DB$2$ and the other half from DB$3$ and DB$4$. The query to each database is symmetric such that for each bit of $A$ that is requested, a bit each from $B$ and $C$ are also requested. All coded pairs of bits from the $3$ messages are requested an equal number of times.
%Also, no specific bit is requested multiple times from any %one database.
Ultimately, the user can decode all  bits of message $A$, because downloaded bits of $B$ and $C$ can be used for decoding (see Table~\ref{table: ex_1}).
In the following, we will first formalize the connection between the FS-PIR and SC-PIR problems and then generalize this example.

%{\color{red} Do we need the following two paragraphs that basically explain the Sun and Jafar scheme?} The query is generated in $3$ rounds and in each round the queries  are symmetry across messages. For instance, in the first round the user requests $1$ bit from each message $A$, $B$ and $C$ from each database. In the second round, for each database, the user requests $3$ coded bits, or $1$ coded bit for each combination of $2$ files. Finally, in the last round the user requests $1$ coded bit from each database for which $1$ bit from each message are coded together.
%
%Moreover, no bit is requested, either alone or in a coded message, more than once from any one database. For instance, bit $b_1^3$ is used for decoding  the transmission from DB$1$; while it is possible to use this bit again in another coded message, it would be clear to DB$1$ that $b_1^3$ is only being used for decoding. In other words, if the user requested $B$, then the user would not request any bit of $B$ twice. The user is able to recover $A$ since all bits of $A$ are included in a transmission and the coded transmissions including bits of $A$ can be decoded as the other coded bits were transmitted by another database. For example, the user can recover $a_2^1$ from DB$3$'s transmission, $a_2^1 + b_2^6$ because it received an uncoded transmission of $b_2^6$ from DB$2$ in the previous round. Similarly, the user can recover $a_1^4$ from DB$2$'s transmission of $a_1^4+b_1^6+c_1^5$ because $b_1^6+c_1^5$ was received from DB$1$ in the previous round.

\subsection{The general connection between the FS-PIR and SC-PIR}
\label{sec: gen_sc1}
%\subsection{Achievable Storage Constrained PIR Scheme when $\frac{N}{t} \in \mathbb{Z}^+$}
%\label{subsec_SC-PIR1}

%\mj{[Is it the case when $\frac{N}{t}$ is an integer?]}
%Define a vector $ \boldsymbol{\alpha} = [ \alpha_1 , \ldots , \alpha_F]\in\Delta_F$
%where $F \in \mathbb{Z}^+$.
Define a vector $ \boldsymbol{\alpha} = [ \alpha_1 , \ldots , \alpha_F]$, %\in\Delta_F$
where $F \in \mathbb{Z}^+$, $\sum_{i=1}^{F} \alpha_i =1$, and $\alpha_f ,\forall f \in [F]$ is rational number such that $\alpha_f L \in \mathbb{Z}^+$.
For all $k \in [K]$, we divide message $W_k$ into $F$ disjoint sub-messages $W_k =  W_{k,1} , \ldots ,  W_{k,F}$ such that for all $f \in [F]$, $|W_{k,f}| = \alpha_f L$ bits.  %sub-message $W_{k,f}$ contains $\alpha_f L$ bits. Next,
For all $f \in [F]$, let
\be
M_f \triangleq \bigcup\limits_{k \in [K]} W_{k,f},
\label{eq_M_f}
\ee
and $\mathcal{N}_f \subseteq [N]$ be a non-empty subset of databases %\mj{[We may want to use a unified notation for sets.]}
which have the sub-messages in $M_f$ locally available to them. The storage contents of database $n \in [N]$ is
\be
Z_n = \left \{ M_f : f \in [F], n \in \mathcal{N}_f \right \},
\label{eq_Z_n}
\ee
where we have the requirement that for any $n \in [N]$,
\be
\sum_{\left \{ f : f \in [F], n \in \mathcal{N}_f\right \} } \alpha_f \leq \mu.
\label{eq_mu_F}
\ee

Given that a user requests file $W_\theta$ for some $\theta \in [K]$, we do the following. For all $f\in [F]$, using a FS-PIR scheme, the user generates a query to privately download $W_{\theta , f}$ from the databases in $\mathcal{N}_f$. In other words, a SC-PIR scheme can be found by applying a FS-PIR scheme to each set of databases $\mathcal{N}_f$. Changing the choice of the FS-PIR scheme or the definitions of $\Nc_f$ will result in new SC-PIR schemes.
%Two examples of such SC-PIR schemes will be provided in the remainder of the paper.

%\if{1}
%\subsection{Achievability}

The rate of the SC-PIR scheme as a function of the rate of the implemented FS-PIR scheme is given in the following theorem.
%described in Section \ref{subsec_SC-PIR1} is given in the following theorem.

\begin{theorem}
\label{th: full2constr}
Given $N,K,F \in \mathbb{Z}^+$ and $\boldsymbol{\alpha}$, %\in \Delta_F$,
split each of the $L$-bit messages $W_1 , \ldots , W_K$  into $F$ sub-messages of size $\alpha_1 L , \ldots , \alpha_F L$ and store them at sets of databases $\mathcal{N}_1 , \ldots , \mathcal{N}_F \subseteq [N]$, respectively.  Given a set of FS-PIR schemes with achievable rates $R_1 , \ldots , R_F$, the achievable rate of privately downloading $W_\theta$, $\theta \in [K]$, from the $N$ storage constrained databases is
  \be
  R = \left(\frac{\alpha_1}{ R_1} + \frac{\alpha_2}{ R_2} + \cdots + \frac{\alpha_F}{ R_F} \right)^{-1}. \label{eq: con_st_rate}
  \ee
\end{theorem}

\begin{IEEEproof}
We first count the number of downloaded bits. For all $f \in [F]$,
%\be
$R_f = \frac{\alpha_f L}{D_f}$
%\ee
where $D_f$ is the number of downloaded bits necessary to privately download $W_{\theta , f}$ of size $\alpha_f L$ bits from the databases in $\mathcal{N}_f$. Therefore, the total number of  bits required to privately download the entirety of $W_\theta$ is
\begin{align}
D &= D_1 + D_2 + \cdots + D_F \nonumber
= L \left( \frac{\alpha_1  }{ R_1} +\frac{\alpha_2 }{ R_2} + \cdots + \frac{\alpha_F }{ R_F}\right).
\end{align}
Since $R = \frac{L}{D}$, we obtain  (\ref{eq: con_st_rate}).
\end{IEEEproof}
%\fi

%\subsection{Adaptation of Sun and Jafar Scheme \cite{sun2017capacity} to Storage Constrained PIR}
%\label{sec: ch1}
\subsection{General Achievable Storage Constrained PIR Scheme When $\frac{N}{t} \in \mathbb{Z}^+$}
\label{subsec_SC-PIR1}
%Given $N$ databases  that each stores a $\mu$ fraction of the library, the question remains of how to divide and distribute messages of the library. Here, we discuss the adaptation of the FS-PIR scheme in \cite{sun2017capacity} to SC-PIR.\footnote{Refer to \cite{sun2017capacity} for specific details on the full storage PIR scheme used here.}

\subsubsection{Storage Placement Scheme}
Given $N \in \mathbb{Z}^+$ and $t \in [N]$ such that $\frac{N}{t} \in \mathbb{Z}^+$, let $F = \frac{N}{t}$ and for each $k \in [K]$, split message $W_k$ into $\frac{N}{t}$ disjoint, equal-size sub-messages, $W_{k,1} , \ldots , W_{k,\frac{N}{t}}$. Furthermore, split the $N$ databases into $\frac{N}{t}$ disjoint groups of size $t$ labeled as $\mathcal{N}_1 , \ldots , \mathcal{N}_{\frac{N}{t}}$. For each $f \in \left[ \frac{N}{t} \right]$, the sub-messages of
\be
M_f = \bigcup\limits_{k\in[K]} W_{k,f}
\ee
are stored at every database of $\mathcal{N}_f$.

\subsubsection{PIR Scheme}
A user desires to privately download message $W_\theta$ for some $\theta \in [K]$. For each $f \in \left[ \frac{N}{t} \right]$, the user generates a query using the scheme of \cite{sun2017capacity}, to privately download $W_{\theta , f}$ from the $t$ databases in $\mathcal{N}_f$. The user combines the downloaded sub-messages, $W_{\theta , 1} , \ldots , W_{\theta , \frac{N}{t}}$ to recover the desired message $W_{\theta}$.

To implement this SC-PIR scheme, each message is split into $\frac{N}{t}$ equal-size, disjoint sub-messages. Furthermore, the adaptation of the FS-PIR scheme of  \cite{sun2017capacity} requires that each sub-message is further split into $t^K$ equal-size, disjoint sub-messages. The resulting SC-PIR requires a total of
   $L = \frac{N}{t} \cdot t^K=   N t^{K-1}$ bits. An example of this SC-PIR scheme  is described  in Section \ref{sec: ex1}.

\subsubsection{Achievable Rate}
The achievable rate of this scheme is summarized as follows.

\begin{theorem}
\label{th: new_cons1}
   Given $N,K,$ and $\mu \in \left[\frac{1}{N},1\right]$, such that $t = \mu N \in [N]$, $\frac{N}{t} \in \mathbb{Z}^+$ and $L =  N t^{K-1}$, for a user to privately download one of $K$ $L$-bit messages from $N$ databases with a storage capacity of $\mu K L$ bits, the achievable rate is
  \be
  R = \left(1 + \frac{1}{t} + \frac{1}{t^2} + \cdots +\frac{1}{t^{K-1}} \right)^{-1}. \label{eq: sc1_rate_1}
  \ee
\hfill$\square$
\end{theorem}

Moreover, it was shown in \cite{attia2018capacity} that (\ref{eq: sc1_rate_1}) is the capacity of SC-PIR for $t\in \mathbb{Z}^+$. While we do not directly prove Theorem~\ref{th: new_cons1} here, in Section \ref{sec: suff_cond} we present a set of sufficient conditions, which this scheme satisfies, for an SC-PIR scheme to meet the capacity.

%\begin{IEEEproof}
%We use the results of Theorem \ref{th: full2constr}. The messages are split into $\frac{N}{t}$ sub-messages each of size $\frac{Lt}{N} = \mu L$ bits. Using the notation from Theorem \ref{th: full2constr}, $\alpha_f = \mu $ for $f \in \left[ \frac{N}{t} \right]$. Furthermore, $F = \frac{N}{t} = \frac{1}{\mu}$ and
%\be
%\sum_{f=1}^{F}\alpha_f = \frac{1}{\mu} \cdot \mu = 1.
%\ee
%The rate to privately download sub-message $W_{\theta , i}$ is defined by the rate of the scheme in \cite{sun2017capacity} to privately download one of $K$ messages from $t$ databases which is
%\be
%R_i = \left(1 + \frac{1}{t} + \frac{1}{t^2} + \cdots +\frac{1}{t^{K-1}} \right)^{-1}.
%\ee
%By Theorem \ref{th: full2constr}, the rate to download the entirety of $W_\theta$ is
%\begin{align}
%R &= \left( \frac{\mu}{R_1} + \cdots + \frac{\mu}{R_{\frac{1}{\mu}}} \right)^{-1}
%= \left( \frac{1}{\mu}\cdot\frac{\mu}{R_1} \right)^{-1}  = R_1 \nonumber \\
%&= \left(1 + \frac{1}{t} + \frac{1}{t^2} + \cdots +\frac{1}{t^{K-1}} \right)^{-1}.
%\end{align}
%\end{IEEEproof}

\section{The proposed SC-PIR Scheme when $\frac{N}{t} \notin \mathbb{Z}^+$}

%\subsection{A New Full Storage PIR Scheme}
%\label{sec-new-FS-PIR}
In Section \ref{sec:connection}, we established a general connection between SC-PIR  and FS-PIR problems. We showed that by properly splitting messages and allocating sub-messages to different groups of databases, a SC-PIR scheme can be derived by applying a separately designed FS-PIR scheme to each group of databases. In particular, when choosing the FS-PIR scheme to be the one in \cite{sun2017capacity}, we obtain a SC-PIR scheme that achieves capacity while requiring  $\frac{N}{t} \in \mathbb{Z}^+$.  In order to remove this restriction, in this section,   we propose a new storage placement and use it in conjunction with the achievable FS-PIR scheme of \cite{sun2017optimal} %that achieve the same optimal rate as the SC-PIR scheme of \cite{attia2018capacity} while requiring $L = Nt^{K-1}$, which is the same as the case when $\frac{N}{t} \in \mathbb{Z}^+$. %less number of bits $L$ per sub-message. This
%Based on the connection between SC-PIR and FS-PIR problems, the proposed new FS-PIR scheme will be applied on the novel storage placement scheme %in Section \ref{sec:new-SC-PIR}
 to obtain a new SC-PIR scheme. This scheme  achieves capacity while requiring only $L =  N t^{K-1}$, which is the same as the scheme of Section \ref{subsec_SC-PIR1} when $\frac{N}{t} \in \mathbb{Z}^+$. %In addition, the proposed new FS-PIR scheme also  achieve the same optimal rate as the SC-PIR scheme of  \cite{sun2017capacity} while only requiring $L=N^{K-1}$ instead of $L=N^K$ in \cite{sun2017capacity}. To illustrate the proposed storage placement and the PIR scheme, we will first present an example and then introduce the general schemes.
% and removing the restriction of $\frac{N}{t} \in \mathbb{Z}^+$. % \cite{sun2017capacity}
%Furthermore,  we will show in Section \ref{sec:new-SC-PIR} that when this new FS-PIR can be applied to derive a new  SC-PIR scheme proposed , in conjunction with the new FS-PIR scheme,  achieves the optimal rate of the state-of-the-art scheme in \cite{attia2018capacity}, while requiring only $L =  N t^{K-1}$ bits per file and removing the restriction of $\frac{N}{t} \in \mathbb{Z}^+$.}
% as required by the SC-PIR scheme described in  Section  \ref{sec: ch1}.
%Later in this section, we demonstrate how this new full-storage scheme can be applied to the storage constrained PIR problem and achieves the optimal rate of the state-of-the-art scheme in \cite{attia2018capacity}. Furthermore, for the storage constrained scheme, we only require $L =  N t^{K-1}$ bits per file and we also alleviate the requirement that $\frac{N}{t} \in \mathbb{Z}^+$ as was the case for the scheme described in Section \ref{sec: ch1}.

\subsection{A Storage Constrained PIR Example when $\frac{N}{t} \notin \mathbb{Z}^+$}

%Here, we apply the new FS-PIR scheme described in Section \ref{sec: sc2} to the SC-PIR problem.
In this example, $N=5$ databases, labeled DB$1$ through DB$5$, collectively store $K=2$ messages, $A$ and $B$, and each has a size of $L=15$ bits. Each database stores an $\mu = \frac{3}{5}$ fraction of the $2$-message library ($t=\mu \cdot N=3$).
\subsubsection{Storage Placement Scheme}
%Specifically, the
%The bits of the messages are labeled such that
Each message is split as follows.
\begin{align}
  A &= \left\{ a_i^j : i \in [5], j \in [3] \right\}, \quad
  B = \left\{ b_i^j : i \in [5], j \in [3] \right\}.
\end{align}
By this labeling, we have essentially split the messages in two phases. The first splitting phase, denoted by the subscript, determines which databases store these bits. The second splitting, denoted by the superscript, is necessary to perform the FS-PIR scheme. For all $f \in [5]$, define
\be
M_f = \bigcup\limits_{j \in [3]}\left( a_f^j \cup b_f^j \right)
\ee
and let the set of databases $\mathcal{N}_f = [-2:0]\oplus_N f$ locally store the bits of $M_f$. \footnote{We impose the following notation: $a \oplus_N b = (a+b-1 \mod N) +1$ and $[a_1:a_2] \oplus_N b =\left\{ a' \oplus_N b : a' \in [a_1 : a_2 ]\right \}$.}
 Note that as opposed to the SC-PIR scheme described in Section \ref{sec: ex1} where the sets of databases $\{\mathcal{N}_f, f=1,\cdots, F\}$ are mutually exclusive, here we allow them to overlap and hence removing the integer constraint of $\frac{N}{t} \in  \mathbb{Z}^+ $.

 As a result, the bits of message $A$ stored at DB $n\in[5]$ are
\be
Z_n = \left\{ a_i^j : i \in \left\{ [0:2] \oplus_N n \right \} , j \in [3]  \right\}.
\ee
Message $B$ is stored among the databases in a similar manner. For instance, DB$2$ stores all bits $a_i^j$ and $b_i^j$ such that $i \in [2:4] $ and DB$5$ stores all bits $a_i^j$ and $b_i^j$ such that $i \in \left\{ 5,1,2 \right\} $.

\begin{table}[h!]
\vspace{-0.4cm}
\normalsize
\centering
{\small
\caption{ Storage Constrained PIR, $N=5$, $K=2$, $\mu = \frac{3}{5}$}
\vspace{-0.2cm}
\label{table: 2}
\begin{tabular}{ |>{\centering}m{1.32cm}|>{\centering}m{1.32cm}|>{\centering}m{1.32cm}|>{\centering}m{1.32cm}|>{\centering}m{1.32cm}| }
 \hline
 DB$1$ & DB$2$ & DB$3$ & DB$4$ & DB$5$ \\
 $( 1,2,3)$ &$( 2,3,4)$ &$( 3,4,5)$ &$( 4,5,1)$ &$( 5,1,2)$ \\
 \hline
 ${\color{red} a_1^3} \quad {\color{red}b_1^2} $ & $a_2^3 \quad b_2^2$ & $a_3^1 \quad b_3^3 $ & $a_4^2 \quad b_4^3 $ & $a_5^2 \quad b_5^1$ \\
 \hline
 $a_2^1+b_2^2$ & $a_3^3+b_3^3$ & $a_4^3+b_4^3$ & $a_5^1+b_5^1$ & ${\color{red} a_1^2+b_1^2}$ \\
 $a_3^2+b_3^3$ & $a_4^1+b_4^3$ & $a_5^3+b_5^1$ & ${\color{red}a_1^1+b_1^2}$ & $a_2^2+b_2^2$ \\
 \hline
\end{tabular}
}
\end{table}
\subsubsection{PIR Scheme}
The queries of a user that desires to privately download message $A$ are shown in Table \ref{table: 2}. The top row of the table  contains  database labels and the $3$-tuple below each database label defines the subscripts of the bits that are locally available to that database. The remaining three rows of the table  show the queries of the user. %Similar to the previous storage constrained PIR example, the user recovers all the bits of $A$ by either receiving the bits or using bits from $B$ for decoding the coded messages.
The user adopts the FS-PIR scheme of \cite{sun2017optimal} to design queries. For instance, to obtain bits $\{a_1^j, j\in[3]\}$, the user applies the FS-PIR to DB$1$, DB$4$, and DB$5$. In the first round, the user obtains $a_1^3$ from DB$1$. In the second round, the user can decode $a_1^1$ from DB$4$'s transmission of $a_1^1 + b_1^2$ because the user had already received $b_1^2$ from the first round transmission of DB$1$ in round $1$. Similarly, the user decodes $a_1^2$ from DB$5$'s transmission of $a_1^2 + b_1^2$.  These transmissions are highlighted in red in Table \ref{table: 2}.
To ensure privacy, the queries are symmetric and no bit is requested more than once from any one database. In this example, $D=20$ bits are downloaded and the rate is $R = \frac{3}{4}$. Comparing to the state-of-the-art SC-PIR scheme of \cite{attia2018capacity}, the rate is the same, but $L$ has been reduced from ${N \choose t}t^K = {5 \choose 3}3^2=90$ to $Nt^{K-1} = 5 \cdot 3^{2-1}= 15$.

\subsection{General Achievable SC-PIR Scheme When $\frac{N}{t} \notin \mathbb{Z}^+$}
\label{sec-new-SC-PIR}

%\subsection{ A General SC-PIR Scheme for arbitrary $t \in [N]$}
\subsubsection{Storage Placement Scheme}
For each $k \in [K]$, message $W_k$ is split into $N$ disjoint equal-size sub-messages $W_{k,1}, \ldots , W_{k,N}$. For all $f\in [N]$, define a set of sub-messages
$M_f = \cup_{k \in [K]}W_{k,f}$
which is locally stored at the set of databases $\mathcal{N}_f = [-(t-1):0]\oplus_N f$.

\subsubsection{PIR Scheme}
A user desires to privately download message $W_\theta$ for some $\theta \in [K]$. For each $f \in \left[ N \right]$, the user generates a query using the scheme of \cite{sun2017optimal}, to privately download $W_{\theta , f}$ from the $t$ databases in $\mathcal{N}_f$. The user combines the downloaded sub-messages, $W_{\theta , 1} , \ldots , W_{\theta , \frac{N}{t}}$ to recover the desired message $W_{\theta}$. Furthermore, if desired, to obtain symmetry across the databases, i.e., each database sends the same amount of coded bit combinations from each file, the user can choose database $f$ to start the query process when privately downloading $W_{\theta , f}$. For more details on the query generation process, see \cite{sun2017optimal}.

%\subsection{Achievability}
\subsubsection{Achievable Rate}
The achievable rate of this SC-PIR scheme is summarized in the following theorem.

\begin{theorem}
\label{th: new_cons2}
 Given $N,K,$ and $\mu \in \left[\frac{1}{N},1\right]$, such that $t = \mu N \in [N]$ and $L =  N t^{K-1}$, for a user to privately download one of $K$ $L$-bit messages from $N$ databases, each with a storage capacity of $\mu K L$ bits, the rate is
\be
R = \left(1 + \frac{1}{t} + \frac{1}{t^2} + \cdots +\frac{1}{t^{K-1}} \right)^{-1}. \label{eq: sc1_rate}
\ee
\end{theorem}

The results of Section \ref{sec: suff_cond} demonstrate that this SC-PIR scheme satisfies the sufficient conditions to meet the capacity.  This proves Theorem \ref{th: new_cons2}.

\section{Sufficient Conditions to Achieve Capacity for SC-PIR}
\label{sec: suff_cond}
%In the literature there are now at least three different capacity achieving SC-PIR schemes: one from \cite{attia2018capacity} and our two newly proposed schemes. It remains an open question as to what are the necessary conditions for a SC-PIR scheme. By using the methodology of Section \ref{sec: gen_sc1}, we essentially show that as long as batches of sub-messages are assigned to groups of $t$ (or $\lfloor t \rfloor$ and $\lceil t \rceil$ for $t\notin \mathbb{Z}^+$) databases, capacity of SC-PIR can be achieved by using capacity achieving FS-PIR schemes. The following theorem summarizes these conditions for both integer and non-integer $t$.

In this section, we provide two sufficient conditions for a storage placement scheme to achieve the SC-PIR capacity.

\begin{theorem}
Given $N,K,F \in \mathbb{Z}^+$ and $\boldsymbol{\alpha}$, %\in \Delta_F$,
split each of the $L$-bit messages $W_1 , \ldots , W_K$  into $F$ sub-messages of size $\alpha_1 L , \ldots , \alpha_F L$ and store them at sets of databases $\mathcal{N}_1 , \ldots , \mathcal{N}_F \subseteq [N]$ according to equations (\ref{eq_M_f})-(\ref{eq_mu_F}). Each database has a storage capacity of $\mu KL$ bits, $\frac{1}{N} \leq \mu \leq 1$, where $t = \mu N \in [1, N]$.
 Assume that a user requests file $W_\theta$ for some $\theta \in [K]$. A SC-PIR scheme is obtained if for all $f\in [F]$,   the user generates a query to privately download $W_{\theta , f}$ from the databases in $\mathcal{N}_f$ using a capacity-achieving FS-PIR scheme.
  % For all $f\in [F]$, using a given capacity-achieving FS-PIR scheme, the user generates a query to privately download $W_{\theta , f}$ from the databases in $\mathcal{N}_f$.
  %Using the methodology of Section \ref{sec: gen_sc1}, to achieve capacity of SC-PIR such that a user privately downloads one of $K$ sub-messages, each of size $L$ bits, from $N$ databases with a storage capacity of $\mu KL$ bits, $\frac{1}{N} \leq \mu \leq 1$, where $t = \mu N$, a capacity achieving FS-PIR scheme is used to download sub-messages and a sub-message placement is defined such that
The resulting SC-PIR scheme is capacity-achieving if the sub-message storage placement satisfies one of the following two conditions:
  \begin{itemize}
    \item[(1)] If $t\in\mathbb{Z}^+$,  $|\mathcal{N}_f| = t$ for all $f \in [F]$
    \item[(2)] If $t \notin \mathbb{Z}^+$,  $|\mathcal{N}_f| \in \{ \lfloor t \rfloor , \lceil t \rceil \}$ for all $f \in [F]$ such that
    \be
    \sum_{f:|\mathcal{N}_f| = \lfloor t \rfloor}\alpha_f=\lceil t \rceil - t
    \ee
     and
     \be
    \sum_{f:|\mathcal{N}_f| = \lceil t \rceil}\alpha_f = t - \lfloor t \rfloor.
    \ee
  \end{itemize}
\end{theorem}
\begin{IEEEproof}
Define $R_{\rm FS}(x)$ as the rate of a capacity achieving FS-PIR scheme to privately download one of $K$ messages from $x$ nodes. Furthermore,
\be
R_{\rm FS}(x) = \left( 1 + \frac{1}{x} + \cdots + \frac{1}{x^{K-1}}  \right)^{-1}
\ee
as was shown in \cite{sun2017capacity}.

For $t \in \mathbb{Z}^+$, it follows from Theorem \ref{th: full2constr} that the rate of the SC-PIR scheme is
\begin{align}
R &= \left( \frac{\alpha_1}{R_{\rm FS}(t)} + \cdots + \frac{\alpha_F}{R_{\rm FS}(t)} \right)^{-1} = R_{\rm FS}(t) \\
&=  \left( 1 + \frac{1}{t} + \cdots + \frac{1}{t^{K-1}}  \right)^{-1} \label{eq: C1}
\end{align}
which  is the capacity of SC-PIR \cite{attia2018capacity}.

For $t \notin \mathbb{Z}^+$, it follows from Theorem \ref{th: full2constr} that
\begin{align}
R &= \left( \frac{1}{R_{\rm FS}(\lfloor t \rfloor)} \sum_{f:|\mathcal{N}_f| = \lfloor t \rfloor}\alpha_f + \frac{1}{R_{\rm FS}(\lceil t \rceil)}\sum_{f:|\mathcal{N}_f| = \lceil t \rceil}\alpha_f \right)^{-1} \\
&= \left( \frac{\lceil t \rceil - t}{R_{\rm FS}(\lfloor t \rfloor)} + \frac{t - \lfloor t \rfloor}{R_{\rm FS}(\lceil t \rceil)} \right)^{-1} \label{eq: C_nonint_t}
\end{align}
and thus
\be
R^{-1} = (\lceil t \rceil - t) R_{\rm FS}^{-1} (\lfloor t \rfloor) + (t - \lfloor t \rfloor) R_{\rm FS}^{-1} (\lceil t \rceil).
\ee
 Note that the point $\left(t,R^{-1}\right)$ is simply an linear interpolation of the two points $\left(\lfloor t \rfloor , R_{\rm FS}^{-1} (\lfloor t \rfloor)\right)$ and $\left(\lceil t \rceil , R_{\rm FS}^{-1} (\lceil t \rceil)\right)$ where the capacity of SC-PIR for $t=x$ is precisely $R_{\rm FS}(x)$. Moreover, it was shown in \cite{attia2018capacity} that the set of achievable points $\left( t, R^{-1} \right)$, is the lower convex hull of the set points $\left \{ \left( t, C_t^{-1} \right) : t \in [N] \right \}$. Therefore, (\ref{eq: C_nonint_t}) meets the SC-PIR capacity.
\end{IEEEproof}

\section{Discussion and Future Work}
Recent works on SC-PIR suggest that coded caching \textit{meets} PIR \cite{attia2018capacity,tandon2018pir}; that is, the file placement solutions of coded caching \cite{maddah2014fundamental} are useful for the SC-PIR sub-message placement problem. In this work, we show that coded caching placement techniques are not necessary for SC-PIR by proposing two novel sub-message placement schemes which achieve the capacity. In the coded caching problem, assigning different files to an exponentially large number of overlapping user groups is necessary to create multicasting opportunities such that a user can cancel ``interference" from a received coded transmission which also serves other users. The SC-PIR problem is less complex in that only one user is being served. In fact, as was demonstrated with our first proposed scheme, it is not necessary for the sub-message placement groups to overlap at all. Moreover, the file (or sub-message) placement paradigms of coded caching and SC-PIR are inherently different. In coded caching, files are being placed among users that wish to download content, while in SC-PIR, sub-messages are being placed among databases which are serving one user's request. Therefore, it is not surprising the two problems could have different solutions for the storage/file placement problem.

The results of Section \ref{sec: suff_cond} show that there exists simple SC-PIR solutions for non-integer $t$. For example, the databases could be split into two disjoint groups, one in which sub-messages are assigned to sub-groups of size $\lfloor t \rfloor$ databases, and another where sub-messages are assigned to sub-groups of size $\lceil t \rceil$ databases. This is contrary to the solution for non-integer $t$ of the coded caching problem where the storage of every user is split into two parts to essentially create two coded caching networks that both span across all users \cite{maddah2014fundamental}. While this coded caching method was proposed to solve the non-integer $t$ SC-PIR problem in \cite{attia2018capacity}, we have shown that this is not necessary.

 This work presents several interesting directions for future work. First, it remains an open problem to determine the minimum  message size $L$ for a given set of SC-PIR parameters. Using a definition of the retrieval rate that is slightly different from that of {\cite{sun2017optimal}, it was shown in \cite{tian2018capacity} that   the minimum $L$ of an FS-PIR problem can be reduced significantly from $N^{K-1}$ in {\cite{sun2017optimal}} to $N-1$. The new FS-PIR scheme   \cite{tian2018capacity} can be readily adapted to our proposed SC-PIR to reduce the message size. Furthermore, the proof techniques therein may be useful to derive the minimum $L$ for a SC-PIR problem.
%Interestingly, the achievable scheme of \cite{tian2018capacity} actually requires that databases store ``dummy" bits in addition to storing all messages. This would have interesting implications if this scheme were applied to the SC-PIR problem, and the exact methodology outlined here could not be used.
Second, another work \cite{wei2018capacity} has considered random placement among  databases  where a database stores a bit of a given message with probability $\mu$. Interestingly, this placement method was also used in \cite{maddah2015decentralized} for the coded caching problem. It will be meaningful to examine alternative random placement strategies for the SC-PIR problem where messages are split into a finite number of sub-messages.

\bibliographystyle{IEEEbib}
\bibliography{references_d2d}

\end{document}